Article Title: **Artificial Intelligence in Assessing Cardiovascular Diseases and Risk Factors via Retinal Fundus Images: A Review of the Last Decade**

Article Category: Advanced Review

Authors:

| |
|---|
| Mirsaeed Abdollahi, MD<br>Affiliation: Nikookari Eye Hospital, Tabriz University of Medical Sciences, Tabriz, Iran.<br>Email: Abdollahi.mirsaeed@gmail.com<br>https://orcid.org/0000-0002-2444-1172 |
| Ali Jafarizadeh, MD, MPH *<br>Affiliation: Nikookari Eye Hospital, Tabriz University of Medical Sciences, Tabriz, Iran.<br>Email 1: Ali.jafarizadeh.md@gmail.com<br>Email 2: Jafarizadeha@tbzmed.ac.ir<br>https://orcid.org/0000-0003-4922-1923 |
| Amirhosein Ghafouri-Asbagh, MD<br>Affiliation: Cardiovascular Research Center, Tabriz University of Medical Sciences, Tabriz, Iran<br>Email: aghafouri98@gmail.com<br>https://orcid.org/0000-0003-2065-1014 |
| Navid Sobhi, MD<br>Affiliation: Nikookari Eye Hospital, Tabriz University of Medical Sciences, Tabriz, Iran.<br>Email: navidsbg.ns1998@gmail.com<br>https://orcid.org/0000-0003-0663-850X |
| Keysan Pourmoghtader, MD<br>Affiliation: Research Center for Evidence-Based Medicine, Iranian EBM Centre: A Joanna Briggs Institute Affiliated Group, Tabriz University of Medical Sciences, Tabriz, Iran<br>Email: Keysanpm96@gmail.com<br>https://orcid.org/0009-0001-2193-2005 |
| Siamak Pedrammehr, PhD<br>Affiliation: Faculty of Design, Tabriz Islamic Art University, Tabriz, Iran<br>Email: s.pedrammehr@gmail.com<br>https://orcid.org/0000-0002-2974-1801 |
| Houshyar Asadi, PhD<br>Affiliation: Institute for Intelligent Systems Research and Innovation (IISRI), Deakin University, VIC 3216, Australia<br>Email: houshyar.asadi@deakin.edu.au |

https://orcid.org/0000-0002-3620-8693

Ru-San Tan, MBBS

Affiliation 1: National Heart Centre Singapore, Singapore.
Affiliation 2: Duke-NUS Medical School, Singapore.

Email: tanrsnhc@gmail.com

https://orcid.org/0000-0003-2086-6517

Roohallah Alizadehsani, PhD *

Affiliation: Institute for Intelligent Systems Research and Innovation (IISRI), Deakin University, VIC 3216, Australia

Email: r.alizadehsani@deakin.edu.au

Postal: 75 Pigdons Rd, Waurn Ponds VIC 3216, Australia
https://orcid.org/0000-0003-0898-5054

U. Rajendra Acharya, PhD

Affiliation 1: School of Mathematics, Physics and Computing, University of Southern Queensland, Springfield, Australia
Affiliation 2: Centre for Health Research, University of Southern Queensland, Australia

Email: Rajendra.Acharya@unisq.edu.au

https://orcid.org/0000-0003-2689-8552

**Conflict of interest**

No author states to have any conflicts of interest.

**Abstract**

**Background:** Cardiovascular diseases (CVDs) are the leading cause of death globally. The use of artificial intelligence (AI) methods—in particular, deep learning (DL)—has been on the rise lately for the analysis of different CVD-related topics. The use of fundus images and optical coherence tomography angiography (OCTA) in the diagnosis of retinal diseases has also been extensively studied. To better understand heart function and anticipate changes based on microvascular characteristics and function, researchers are currently exploring the integration of AI with non-invasive retinal scanning. There is great potential to reduce the number of cardiovascular events and the financial strain on healthcare systems by utilizing AI-assisted early detection and prediction of cardiovascular diseases on a large scale.

**Method:** A comprehensive search was conducted across various databases, including PubMed, Medline, Google Scholar, Scopus, Web of Sciences, IEEE Xplore, and ACM Digital Library, using specific keywords related to cardiovascular diseases and artificial intelligence.

**Results:** The study included 87 English-language publications selected for relevance, and additional references were considered. This paper provides an overview of the recent developments and difficulties in using artificial intelligence and retinal imaging to diagnose cardiovascular diseases. It provides insights for further exploration in this field.

**Conclusion:** Researchers are trying to develop precise disease prognosis patterns in response to the aging population and the growing global burden of CVD. AI and deep learning are revolutionizing healthcare by potentially diagnosing multiple CVDs from a single retinal image. However, swifter adoption of these technologies in healthcare systems is required.

**Graphical/visual abstract and caption**

Exploring the Landscape of AI in Cardiovascular Risk Assessment: Comparative Performance of CNN, RNN, and ANN Models, Integrated Pathways for AI Implementation, and Addressing Challenges in CVD Risk Assessment with Artificial Intelligence

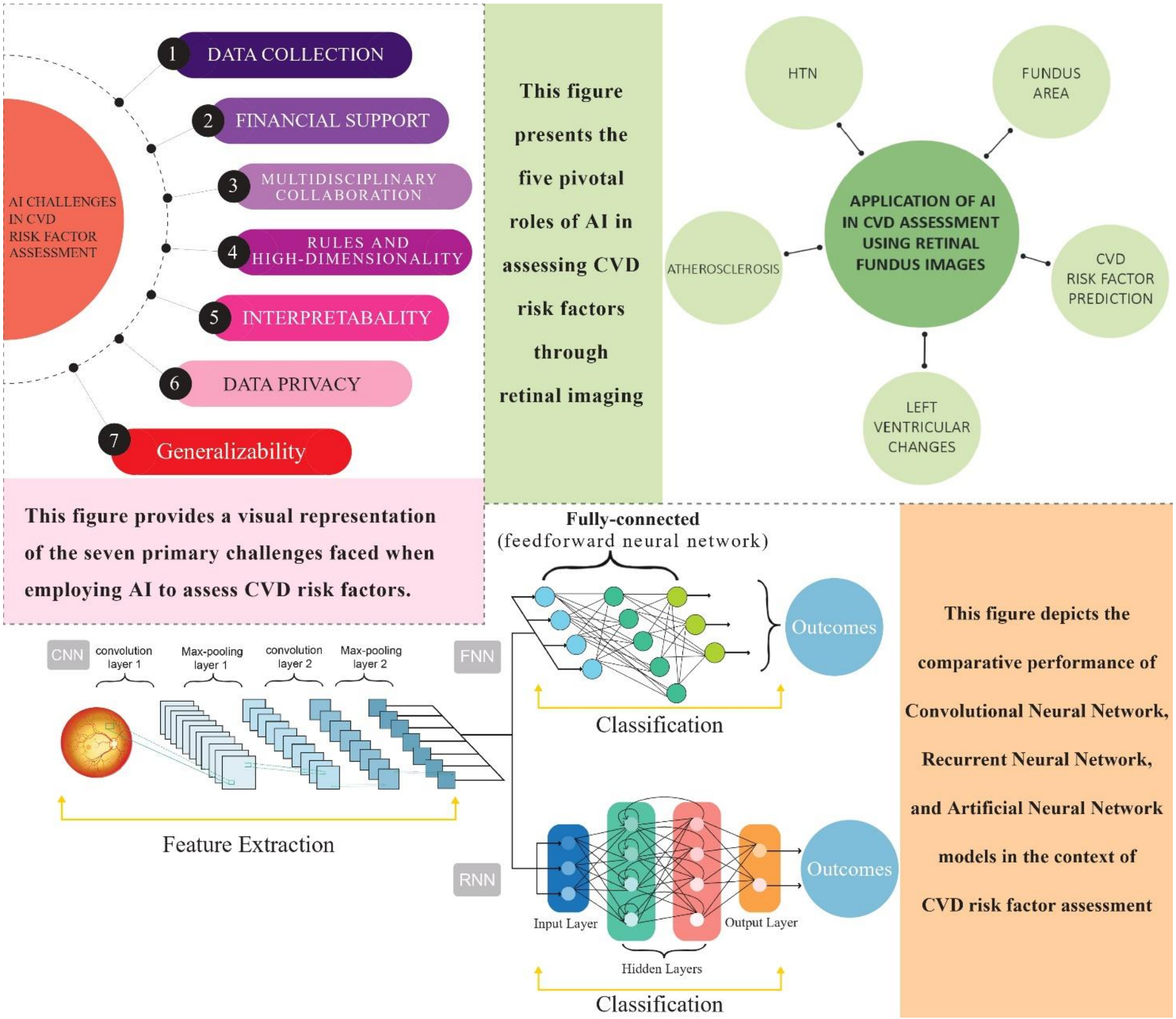

## 1. Introduction

Cardiovascular diseases (CVDs) claimed the lives of over 18.6 million people in 2019, as opposed to 17.9 million in 2016 (Liu et al., 2022). These figures, which are more common in wealthy nations, account for 32% of all deaths and 40% of deaths among people under the age of 70 (Al-Absi et al., 2022; Organization, 2017). In 2015, the economic impact of CVD in the United States was more than $351.3 billion (Virani et al., 2020b). Various risk factors like age, gender, smoking, obesity, diabetes, high blood pressure, high cholesterol levels, and lack of activity are risk factors for the development of CVDs (Ghantous et al., 2020). Diagnostic tools like magnetic resonance imaging (MRI), echocardiography, computed tomography, single-photon emission computerized tomography, and ultrasound play roles in identifying CVD cases (Seetharam & Min, 2020). Retinal imaging is becoming popular because it is non-invasive and can directly visualize the retina and its blood vessels. Some studies indicate a correlation between retinal vascular characteristics' changes and CVD (Farrah et al., 2020; Wang et al., 2019). However, analyzing retinal images requires expertise and could be skewed by personal prejudices. Artificial intelligence (AI) has the potential to expedite image analysis by identifying patterns associated with medical conditions. Moreover, retinal imaging provides a budget option that aligns well with AI-based analysis. (Mayya et al., 2023).

AI tools include machine learning (ML), which predicts datasets using features crafted by domain experts, and deep learning (DL), which enables end-to-end learning on large annotated datasets without the need for human intervention (Bazargani et al., 2024; Tsoi et al., 2021). DL, on the other hand, is a subset of ML that focuses on multilayer neural networks. In contrast, ML includes supervised, unsupervised, and reinforcement learning methods that typically make use of structured data and look for patterns and relationships within datasets (Farabi Maleki et al., 2023). ML typically works with structured data, aiming to recognize patterns and relationships within datasets. In contrast, DL is particularly skilled at managing unstructured data and has exhibited impressive performance in tasks such as speech recognition, computer vision, and natural language processing. ML approaches are often employed for solving simpler problems that can be executed on standard computers, whereas DL models require substantial amounts of data for training and sophisticated algorithms to iteratively refine their performance over time (Alzubaidi et al., 2021;

Ashayeri et al., 2024). There are many different types of DL networks, each designed to fulfill a particular purpose. Feedforward neural networks (FNNs) are perfect for tasks like classification because information flows in a single direction from the input layer through the hidden layers and out to the output layer. Because recurrent neural networks (RNNs) can retain information from previous inputs, they are well-suited for processing sequential data, including time series, natural language processing, and speech recognition. For specialized tasks like image recognition and medical image analysis, convolutional neural networks (CNNs) use spatial hierarchies in data (Chauhan & Singh, 2018; Schmidhuber, 2015; Yousefi et al., 2024). Different CNN architectures, such as Xception, U-Net, DenseNet, residual networks (ResNet), GoogLeNet/Inception, Visual Geometry Group (VGG) network, and AlexNet, have developed to address particular issues (Dhillon & Verma, 2020; Hosseini et al., 2020).

This study reviewed the AI architectures and techniques used to analyze retinal images and predict the presence of CVD and its risk factors. The results, importance, and application of these methods are discussed along with their current and future prospects and challenges..

## 2. Material and methods

We performed a comprehensive review of the existing literature using multiple academic databases, such as PubMed, Medline, Google Scholar, Scopus, Web of Sciences, and IEEE Xplore. The scope of the investigation was restricted to scholarly publications published exclusively in English up to July 1, 2023. By reviewing the reference sections of the screened papers, more articles were gathered. For this comprehensive review, an elaborate search strategy was built using various combinations of the following keywords: "cardiovascular diseases," “cardiovascular abnormality,” "heart," "retina," “fundus image,” "retinal image," "diagnosis," “prognosis,” "risk factor," "artificial intelligence," "deep learning," "machine learning," “prediction,” “neural network” and "machine vision".

## 3. Results

The initial search identified 2056 research papers. 1219 duplicate papers were removed. A review of the titles and abstracts was then conducted by study authors [MA, AJ, AGA, NS], which led to the exclusion of 698 articles that did not meet the specified criteria. A comprehensive review of the full-text articles was conducted by study authors [MA, AJ, AGA, NS], which excluded another 52 publications deemed unsuitable. A final total of 87 papers were reviewed in this study (Figure 1).

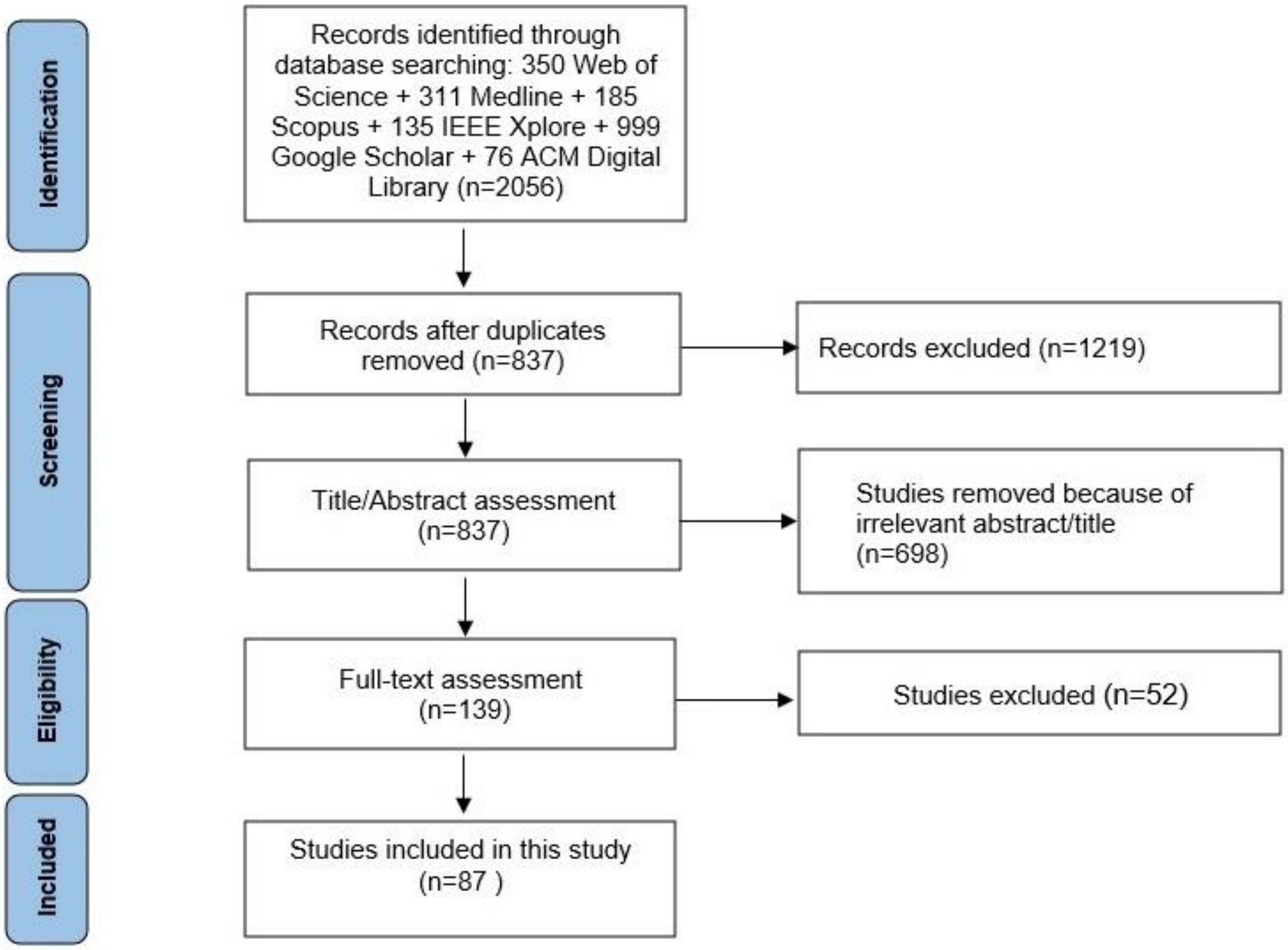

Figure 1 The flowchart of the search and review process.

## 4. Discussion

### 4.1 Predicting CVD Risk Factors from Fundal Retinal Images Using AI

Finding those without CVD who are most at risk of getting CVD is a major duty of preventive population health programs (Weintraub et al., 2011). Framingham risk score (Wilson et al., 1998), pooled cohort equation (PCE), QRESEARCH risk estimator version 3 (QRISK3), systematic coronary risk evaluation-2 (SCORE2) ("SCORE2 risk prediction algorithms: new models to estimate 10-year risk of cardiovascular disease in Europe," 2021), and atherosclerotic cardiovascular disease (ASCVD) risk estimator plus (American College of Cardiology. ASCVD Risk Estimator; American College of Cardiology: Washington, DC, USA, 2022., 2022) are established risk measures for estimating the 10-year risk of developing CVD, i.e., experiencing cardiovascular events or fatalities. The accuracy of the risk-stratification algorithms would probably be restricted to populations in which they have been developed and require external validation if they were used for other populations. The scores can only be produced by typical input of multiple clinical parameters and laboratory results, such as age, sex, ethnicity, blood pressure, smoking status, blood glucose, and cholesterol levels (Arnett et al., 2019; Francula-Zaninovic & Nola, 2018).

Since the fundus is the only place in the body where blood vessels (retinal arteries) can be directly seen, it offers a unique window into vascular health. It is an attractive idea to screen the population for CVD risk with a single non-invasive "eye check." The general public is more likely to participate in eye screening than medical visits for cardiovascular assessment (Siegfried K Wagner et al., 2020). Between 2009 and 2013 in the United Kingdom (UK), only 12.8% of adults over 40 underwent CVD screening (Robson et al., 2016), compared to over 50% who underwent routine eye exams (The State of the National Eye Health 2017: A Year in Review. RNIB (2018). , 2018; S. K. Wagner et al., 2020)

Fundal examination is traditionally conducted by doctors using either ophthalmoscopes or slit lamps. However, digital fundal photography has emerged as an essential tool for screening eye diseases. According to Sinhababu et al. (Sinhababu, 2022), there are noteworthy associations between fundal retinal images and both CVD and CVD risks. Retinal images require expertise and time to interpret manually but are, fortunately, amenable to AI-assisted analysis. Arnould et al. (Arnould et al., 2021) used fundal images of 144 patients who did not have known macular degeneration, diabetic retinopathy, or vascular occlusive disease to predict cardiovascular parameters using supervised ML. Inputting microvascular network geometrical features extracted from fundal photographs, as well as optical coherence tomography angiography, using the Singapore "I" Vessel Assessment software to naïve Bayes classification algorithm, the model was able to predict CVD risks with varying degrees of accuracy: American Heart Association (AHA) risk score, 81.25%; SCORE, 75.64%; and Syntax score, 96.53%. Poplin et al. (Poplin et al., 2018) used an Inception-v3-based DL model to evaluate 48,101 and 236,234 fundus images from UK Biobank and the open-access EyePACS diabetic retinopathy dataset, respectively. They showed that age (accuracy 78%, ± 5 years error margin) and sex (area-under-curve [AUC] 0.97) could be accurately predicted from retinal images alone. For the prediction of glycated hemoglobin (HbA1c) and index of blood sugar control in diabetic patients, the accuracy was 79% (± 2% error margin) with a low coefficient of determination ($R^2$) of 0.09. A potential explanation for the latter is the higher average HbA1c (with a wider distribution) and age, which attenuates the accuracy of the HbA1c prediction in the EyePACs population. For the prediction of smoking status, the AUC was 0.71. Kim et al. (Kim et al., 2020) studied 412,026 fundus images and health data (hypertension, diabetes, and smoking status) of 155,449 subjects who visited a South Korean health center between 2003 and 2016 and developed a ResNet-152-based DL predictive model. Model accuracies for age prediction (± 5 years error margin) were 82.8%, 77.6%, 77.0%, and 85.6% in the normal (without hypertension, diabetes, or smoking), hypertension, diabetes, and smoker groups, respectively, and for age prediction (±3 years error margin), >55% in all groups. Age prediction was the highest in the third and fourth decades without significant sex differences; above 60 years of age, the accuracy of age prediction gradually decreased. Nusinovici et al. (Nusinovici et al., 2022) trained a DL algorithm with multilayer VGG architecture on 129,236 fundus images from 40,480 participants in the "RetiAGE" Korean Health Screening study to identify age-related retinal patterns without using any other information. The achieved area under the receiver operating characteristic curve of 0.968 (95% CI 0.96 to 0.97) and area under the precision-recall curve of 0.83 (95% CI 0.83 to 0.84) for correctly identifying individuals aged 65 years and older. Gerrits et al. (Gerrits et al., 2020) trained their MobileNet-V2-based model on 3,000 Qatar Biobank retinal fundus images and made predictions for age and sex using statistical measures. For age prediction, the model attained a high $R^2$ value of 0.85 (95% CI 0.83 to 0.87) with a mean absolute error of 3.21 years. They obtained an AUC of 0.96 (95% CI 0.95 to 0.97) and a high accuracy of 0.90 (95% CI 0.89 to 0.91) for sex prediction. For HbA1c prediction, the $R^2$ value was lower at 0.34. For smoking status prediction, the AUC was 0.78. The performance and generalizability of the model could be affected by variations in dataset size, demographics, image quality, and model architecture, according to the authors.

Zhang et al. (Zhang et al., 2020) developed transfer learning with the Inception-v3 CNN model to predict hyperglycemia and dyslipidemia from fundus retinal images of 625 patients with chronic diseases in China. After being trained on 1222 retinal images of superior quality and more than 50 measurements of biochemical and anthropometric parameters, the model attained 78.7% accuracy and 0.880 AUC for detecting fasting hyperglycemia, 66.7% accuracy, and 0.703 AUC for detecting hypertriglyceridemia. Zekavat et al. (Zekavat et al., 2022) developed a U-Net architecture to analyze retinal vasculature in relation to phenome-wide and genome-wide studies. The model was trained on 97,895 fundus retinal images of 54,813 participants from the UK Biobank. They observed that one standard deviation increments in glucose and HbA1c levels were linked to decreases in retinal vascular density ($\beta_{Glucose}$, -0.03; $\beta_{HbA1c}$, -0.01) and fractals dimensions ($\beta_{Glucose}$, -0.02; $\beta_{HbA1c}$, -0.02), which signified structural vascular deteriorations. Vaghefi et al. (Vaghefi et al., 2019) analyzed 165,104 fundus retinal images from the Auckland Diabetic Eye Screening Database using a CNN-based DL model with five convolution layers, five pooling layers, and three fully-connected layers. The model predicted smoking status with an AUC of 0.86, which outperformed Gerrits et al. (Gerrits et al., 2020), in which smoking status was assessed differently, and Poplin et al. (Poplin et al., 2018). Dai et al. (Dai et al., 2020) developed a U-Net architecture for segmenting retinal blood vessels, which was trained on the Digital Retinal Images for Vessel Extraction dataset comprising 2012 fundus images sourced from 1007 and 1005 hypertensive and normotensive subjects, respectively, in China. A concise CNN structure was employed for training and classification that contained five convolution layers, five pooling layers, and two fully connected layers. The model predicted hypertension with an AUC of 0.65.

Beyond the conventional cardiovascular risk factors predictions, the AI model can also predict other CVD-related parameters using fundus retinal images, e.g., body mass index (57% accuracy, ± 3 kg/$m^2$ error margin) in Poplin et al. (Poplin et al., 2018) and drinking status, salty taste, body mass index (BMI), waist-to-hip ratio, and hematocrit with area-under-curve (AUCs) >0.70 in Zhang et al. (Zhang et al., 2020). Table 1 shows recent advances in CVD risk prediction.

**Table 1** Recent Advancements in CVD Risk Prediction.

| Study | Dataset | Training method | Results |
|---|---|---|---|
| **Arnould e**t al. **2021 (Arnould et al., 2021)** | 144 patients; microvascular network geometrical features extracted using Singapore "I" Vessel Assessment Software | Supervised ML; classification (k-nearest neighbors, discriminant analysis, and naïve Bayes) and regression (decision trees). | Predicted AHA, SCORE, and Syntax risk scores with 81.25%, 75.64%, and 96.53% accuracy, respectively. |
| **Poplin et al. 2018 (Poplin et al., 2018)** | 284,335 subjects; pooled UK Biobank and EyePACS datasets | Inception-v3, deep neural network | Predicted age (78% accuracy), body mass index (57% accuracy), HbA1c (79% accuracy), sex (AUC 0.97), smoking status (AUC 0.71), |
| | | | and major adverse cardiovascular events within five years (AUC 0.70). |
| **Kim et al. 2020 (Kim et al., 2020)** | 155,449 patients. | CNNs-based on ResNet-152, transfer learning using pre-trained weights from ImageNet, specific loss functions, and optimization schemes. | Predicted age (82.8% accuracy, normal group; 77.6%, hypertension group; 77%, diabetes group; 85.6%, smoker group; >55%, all groups). |
| **Nusinovici et al. 2022 (Nusinovici et al., 2022)** | 40,480 subjects; Korean health screening study, "RetiAGE" | DL-based on multilayer VGG architecture. | Predicted age > 65 years (AUROC 0.968, 95% CI: 0.96 – 0.97; AUPRC 0.83, 95% CI: 0.83 – 0.84). |
| **Gerrits et al. 2020 (Gerrits et al., 2020)** | 3,000 patients; fundus images from Qatar Biobank (data augmentation was applied, including horizontal flipping, random rotation, and random shifting) | DL-based on MobileNet-V2 architecture. | Predicted age ($R^2$ = 0.85), sex, and smoking status prediction with an accuracy of 0.90 and 0.81, respectively. |
| **Zhang et al. 2020 (Zhang et al., 2020)** | 625 patients | Transfer Learning with the Inception-v3 CNN model. The training process involved retraining the fully-connected and SoftMax layers while keeping the convolutional layers of Inception-v3 fixed | The accuracy in detecting hyperglycemia and dyslipidemia was reported at 78.7% and 66.7%, respectively. |
| **Zekavat et al. 2022 (Zekavat et al., 2022)** | 97,895 retinal fundus images from 54,813 UK Biobank participants | DL based on the U-Net architecture | one standard deviation increments in glucose and HbA1c levels were linked to decreases in retinal vascular density ($\beta_{Glucose}$, -0.03; $\beta_{HbA1c}$, -0.01) and fractals dimensions ($\beta_{Glucose}$, -0.02; $\beta_{HbA1c}$, -0.02). |
| **Vaghefi et al. 2019 (Vaghefi et al., 2019)** | 165,104 retinal fundus images were utilized from the Auckland Diabetic Eye Screening Database | DL uses the CNN model with five convolution layers, five pooling layers, and three fully-connected layers. | The model predicted smoking status with an AUC of 0.86 |
| **Dai et al. 2020 (Dai et al., 2020)** | 2012 fundus images from 1007 hypertensive and 1005 normotensive subjects of the Digital Retinal Images for Vessel Extraction dataset | With CNNs based on U-Net-based segmentation of retinal vessels | The model predicted hypertension with an AUC of 0.65. |

**4.2 Application of AI in fundus images.**

More helpful information about specific CVD and risk factors can be found in specific sections of the fundus (Figure 2). Researchers employ a DL technique called soft attention to locate anatomical regions the model uses to make predictions, i.e., a heatmap of the most predictive pixels within the image. Vaghefi et al. (Vaghefi et al., 2019) demonstrated that in the attention maps from contrast-enhanced fundus images, the retinal vessels, perivascular region, and fovea consistently provided the most information for predicting an individual's smoking status. In Poplin et al. (Poplin et al., 2018), the predictive models designed to estimate HbA1c levels generally focused on perivascular regions, whereas models for sex prediction primarily concentrated on the optic disc, vessels, and macula. For other predictions, including systolic blood pressure and BMI, the attention was less specific, manifesting as either uniform attention across the image or focusing on the image's border, suggesting that the indicators for these particular predictions were more broadly dispersed. In contrast, in Dai et al. (Dai et al., 2020), arterial/venous bifurcation regions were highlighted in the heat maps generated by their model for predicting hypertension. The observed discrepancy could be ascribed to ethnic variations within the study populations. Kim et al. (Kim et al., 2020) demonstrated the significance of the fovea and retinal vessels in sex determination in an experiment. They first created modified retinal fundus images in which the fovea or retinal vessels were removed. Then they observed that the AUCs for sex prediction were 0.881 (95% CI: 0.877 - 0.885) and 0.682 (95% CI: 0.676 - 0.688) for images without the fovea and without the retinal vessels, respectively, compared to AUC of 0.97 (95% CI: 0.96 - 0.98) for the original images. Additionally, the model's class activation maps highlighted varying activation levels in different areas, such as the fovea, optic disc, and retinal artery, with significant activation in the region near the blood vessels among female subjects.

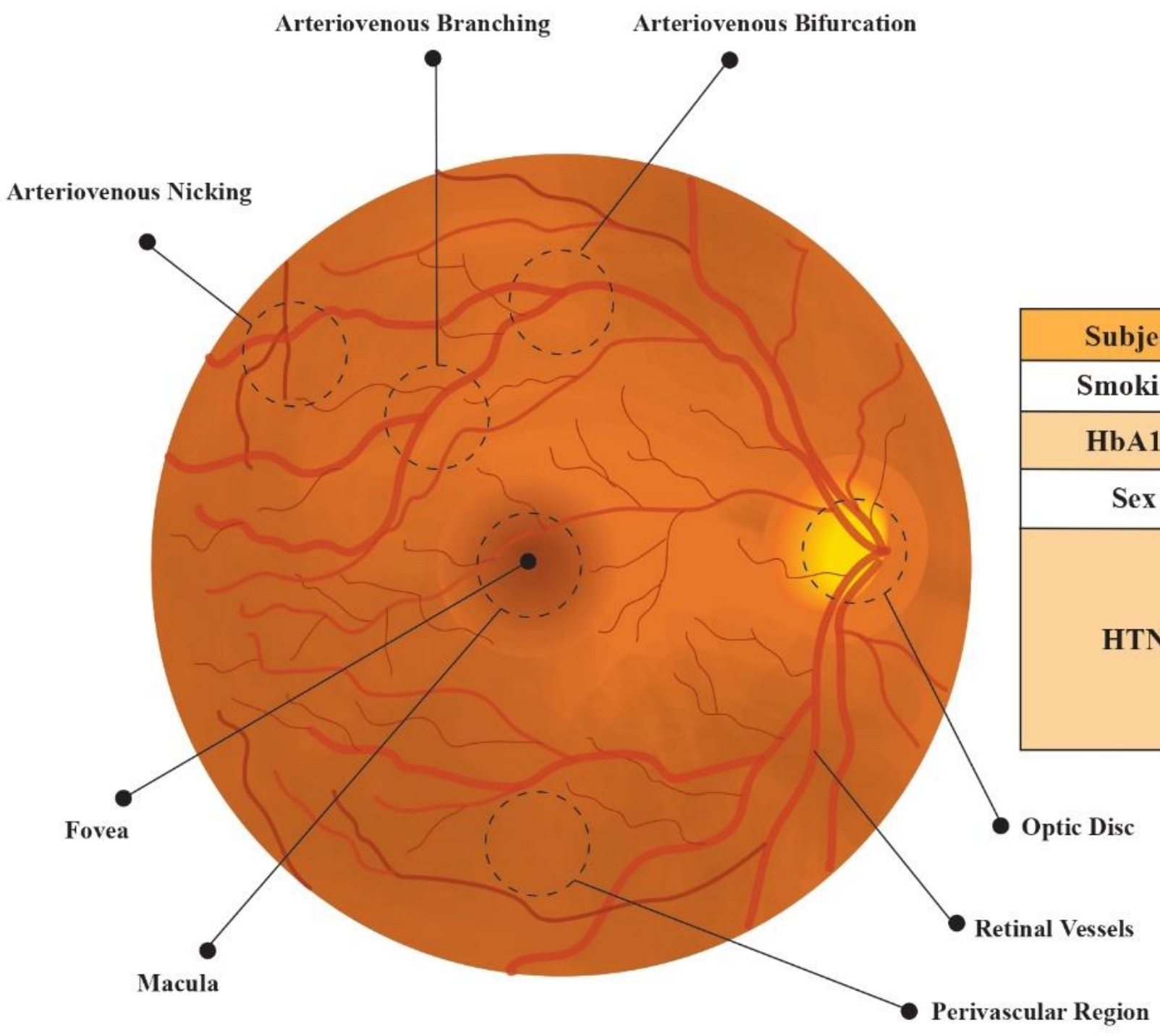

| Subject | Site |
|---|---|
| Smoking | Vessels, Perivascular Region, Fovea |
| HbA1C | Perivascular Region |
| Sex | Optic Disc, Macula |
| HTN | Arteriovenous Bifurcation<br>Arteriovenous Branching<br>Arteriovenous Nicking |

**Figure 2** - Schematic fundoscopic image of the retina showing the optic disc, fovea, macula, retinal vessels, and perivascular regions. Specific regions predictive of CVD risk factors are listed. HTN: hypertension.

### 4.3 Application of AI in predicting hypertension with fundus images

Hypertension and its associated complications are responsible for more than half of the mortality due to CVDs (Virani et al., 2020a). Over time, it causes structural and functional damage to blood vessels (Houben et al., 2017), altering small arteries and microcirculation, leading to end-organ damage in various systems such as the heart, kidneys, brain, eyes, and peripheral vessels (HA et al., 1992; Rizzoni et al., 2003). Hypertension-induced retinal microcirculatory abnormalities include generalized or focal narrowing of retinal arterioles, arteriovenous nicking, retinal infarction (see as white “cotton wool spots”), retinal hemorrhage, microaneurysms, as well as altered retinal geometry: tortuosity, fractal dimension, and vessel branching angle (Smith et al., 2004; Wang et al., 2008; Wong et al., 2004). By assessing these alterations in the retinal microcirculation, we are able to predict the presence of systemic hypertension.

Poplin et al. (Poplin et al., 2018) reported accuracies of 72% (±15 mmHg error margin) and 79% (±10 mmHg error margin) for model prediction of systolic and diastolic blood pressures, respectively, from the fundus images of 12,026 participants in the UK biobank dataset. There was a linear correlation between the predicted and actual systolic blood pressures as the latter increased to approximately 150 mmHg. In Dai et al. (Dai et al., 2020), fundus retinal images were divided into segmented (blood vessels) groups and enhanced groups. The model had an AUC of 0.65 and an accuracy of 60.94% in predicting hypertension. In another study on Chinese patients, Zhang et al. (Zhang et al., 2020) reported 68.8% accuracy and 0.766 AUC for predicting hypertension via their transfer learning Inception-v3-based DL model.

### 4.4 Atherosclerosis in retinal vessels and coronary artery diseases

Systemic arteriosclerosis, which is typically caused by the buildup of atherosclerotic cholesterol plaque in the arterial endothelium, frequently affects the retinal arteries. Wall thickening, lumen narrowing, and ultimately occlusion are the characteristics of this condition(Tien Yin Wong et al., 2005). These alterations mirror systemic diseases like hypertension and hyperlipidemia, as well as the severity, onset (acute or chronic), and duration, which can influence the extent and degree of vascular changes. Hypertension not only causes retinal arteriosclerosis but can also induce vasoconstriction. Population studies have demonstrated that retinal vasculature perturbations, ranging from relatively subtle retinal arteriolar constriction to overt retinal vascular damage, are indicators of preclinical CVD and are predictive of major adverse severe cardiovascular outcomes events and mortality (Seidelmann et al., 2016; Wong et al., 2008).

Gao et al. (Gao et al., 2021) created a four-channel CNN-based DL model with contour augmentation and image merging, which resulted in an accuracy of 65.354% for grading arteriosclerotic retinopathy. Chang et al. (Chang et al., 2020) developed a DL funduscopic atherosclerosis score (DL-FAS) by training their DL model on 15,408 fundus retinal images of patients with or without atherosclerosis, as confirmed by carotid artery sonography. The model was trained using the fundus retinal images to predict the scores for carotid atherosclerosis. Optimizing the prediction also involved adjusting the model's weights. In this retrospective cohort of 32,227 participants aged 30 to 80 years, the model achieved 0.583, 0.891, and 0.404 accuracy, sensitivity, and specificity, respectively, for predicting carotid atherosclerosis. The participants were followed up for a median of 7.6 years. DL-FAS outperformed the established Framingham risk score for the prediction of cardiovascular mortality.

A preclinical indicator of coronary atherosclerosis, the coronary artery calcium score measured during coronary artery computed tomography is highly correlated with the likelihood of developing CVD and is widely used to stratify cardiovascular risk in conjunction with clinical risk factors (Goff et al., 2014). Retinal images are used by Rim et al.'s DL algorithm, RetiCAC (Rim et al., 2021), to predict coronary artery calcium scores. Their model outperformed models based on clinical parameters, achieving an AUC of 0.742 (95% CI: 0.732–0.753) for binary classification of the presence of coronary artery calcium using only retinal images. 216,152 fundus photos from five datasets obtained from facilities in Singapore, South Korea, and the UK were used to train the model. Son et al. (Son et al., 2020) developed an Inception-v3 architecture for predicting coronary artery calcium based on retinal fundus images. Bilateral fundus photography and a same-day coronary artery calcium assessment were performed on the study participants. The AUCs for identifying coronary artery calcium scores higher than 100 and absent coronary calcium using both unilateral and bilateral fundus images were 83.2% and 82.3%, respectively. The authors also improved the performance of their model by incorporating clinical risk factors, such as age, gender, and hypertension, into their DL algorithm in addition to other fundal image inputs, such as fovea-inpainted, vessels-inpainted, unilateral, and bilateral fundus images. To classify coronary artery calcium scores as either below or above 400, Barriada et al. (Barriada et al., 2022) created a hybrid model by fusing VGG16 and conventional ML with fundus images and clinical data (such as age and the presence of diabetic retinopathy). From the combined analysis of fundal image and clinical data, they proposed two useful applications: the first application, which was for clinical diagnosis, produced a recall of 75%, and the second application, which was for image-based diagnosis, produced a precision of 91%.

Retinal vessel caliber has been associated with cardiovascular conditions such as stroke, coronary heart disease, aortic stiffness, and carotid atherosclerosis (Wong et al., 2001; Tien Y Wong et al., 2005). Cheung et al. (Cheung et al., 2021) trained a DL model to measure the caliber of retinal arteriolar and venular on normalized and cropped fundus images derived from a dataset comprising more than 70,000 samples. The connected layer of their DL model outputs vessel caliber measurements of input images.

There was strong consensus on retinal vessel caliber measurements between the models and human experts, with correlation values ranging from 0.82 to 0.95. On a prospective dataset, retinal vessel caliber measurements were associated with CVD risk variables like systolic blood pressure, body mass index, cholesterol, and glycated hemoglobin levels.

### 4.5 Fundus Images in Predicting Left Ventricular Changes

Left ventricle (LV) remodeling is the term used to describe the change in the shape and function of the LV due to molecular, cellular, and interstitial structural changes caused by myocardial injury, such as myocardial infarction or hemodynamic overload (Opie et al., 2006; Udelson & Konstam, 2011; Zeng et al., 2017). These changes have the potential to cause left ventricular hypertrophy as well as heart failure (HF) (Heidenreich et al., 2022), and arrhythmias such as atrial fibrillation and potentially fatal ventricular tachycardia and fibrillation. (Ravassa et al., 2017; Tsai et al., 2021).

Studies have shown that retinal microvascular changes, such as retinal arteriolar narrowing or retinal venular widening, are associated with subclinical macrovascular disease, which shares the same pathophysiological pathways with and may be linked to LV remodeling, LV hypertrophy, and HF (Chaikijurajai et al., 2022; Cheung et al., 2007; Gupta & Reddy, 2021; Tikellis et al., 2008; Tien Y Wong et al., 2005). Cheung et al. (Cheung et al., 2007) demonstrated that concentric LV remodeling was linked to narrower retinal arterioles in a study involving 4,593 patients who had no prior cardiac disease (OR: 2.06; 95% CI: 1.57 - 2.70). Hondur et al. (Hondur et al., 2020) showed that congestive HF causes structural changes in the choroid by altering the blood flow, with slow blood flow in large vessels causing venous blood pooling in the Haller layer and subsequent compression of the inner choroid. Chandra et al. (Chandra et al., 2019) found correlations between decreased central retinal artery equivalent and increased central retinal vein equivalent with echocardiography-assessed left ventricular systolic and diastolic dysfunction in a prospective cohort study involving four medical centers in the United States. There is an unmet clinical need because no publication has been made on using AI to predict HF or LV alterations using fundus images.

Table 2 summarizes the performance metrics of various models and architectures employed to detect cardiovascular disease and risk factors. Accuracy, area under the curve (AUC), R-squared values, and P-values from various studies are among the metrics.

**Table 2** Performance metrics of models and architectures for cardiovascular disease and risk factors.

| Study (year) | AI model | Architecture | Investigation parameters | Results |
|---|---|---|---|---|
| **CVD's risk factors** | | | | |
| **Arnould et al. (2021) (Arnould et al., 2021)** | Supervised ML | k-nearest neighbors, discriminant analysis, and naïve Bayes | Predicted AHA, SCORE, and Syntax risk scores | 81.25%, 75.64%, and 96.53% accuracy, respectively. |
| **Poplin et al. (2018) (Poplin et al., 2018)** | Deep neural network | Inception-v3 | Age | 78% accuracy |
| | | | Body mass index | 57% accuracy |
| | | | Sex | AUC 0.97 |
| | | | Smoking status | AUC 0.71 |
| | | | Major adverse cardiovascular events within five years | AUC 0.70 |
| | | | HbA1c | 79% accuracy |
| **Kim et al. (2020) (Kim et al., 2020)** | DL & transfer learning (using pre-trained weights from ImageNet) & specific loss functions and optimization schemes. | CNNs based on ResNet-152 | Age | normal group=82.8%, hypertensive group=77.6%, diabetes group=77%, smoker group=85.6%, all groups >55% accuracy. |
| **Nusinovici et al. (2022) (Nusinovici et al., 2022)** | DL | Multilayer VGG | Predicted age > 65 years | AUROC 0.968, 95% CI: 0.96 – 0.97; AUPRC 0.83, 95% CI: 0.83 – 0.84 |
| **Gerrits et al. (2020) (Gerrits et al., 2020)** | DL | MobileNet-V2 | Age prediction | $R^2$ = 0.85 (95% CI 0.83 to 0.87) |
| | | | sex prediction | $R^2$ = 0.90 (95% CI 0.89 to 0.91) |
| | | | | AUC=0.96 (95% CI 0.95 to 0.97) |
| | | | HbA1c prediction | $R^2$ = 0.34 |
| | | | smoking status prediction | AUC=0.78 |
| **Zhang et al. (2020) (Zhang et al., 2020)** | Transfer Learning and DL | Inception-v3 | Hyperglycemia detection | 78.7% accuracy and AUC of 0.880 |
| | | | Dyslipidemia detection | 66.7% accuracy and AUC of 0.703 |
| **Vaghefi et al. (2019) (Vaghefi et al., 2019)** | CNN | – | smoking status prediction | AUC of 0.86 |
| **Dai et al. (2020) (Dai et al., 2020)** | CNN | U-Net | Hypertension prediction | AUC of 0.65 |
| **Fundus Area** | | | | |
| **Vaghefi et al. (2019) (Vaghefi et al., 2019)** | CNN | – | smoking status prediction | AUC of 0.86 |
| **Poplin et al. (2018) (Poplin et al., 2018)** | Deep neural network | Inception-v3 | Age | 78% accuracy |
| | | | Body mass index | 57% accuracy |
| | | | Sex | AUC 0.97 |
| | | | Smoking status | AUC 0.71 |
| | | | Major adverse cardiovascular events within five years | AUC 0.70 |
| **Dai et al. (2020) (Dai et al., 2020)** | CNN | U-Net | Hypertension prediction | Accuracy of 60.94% and AUC of 0.65 |
| **Kim et al. (2020) (Kim et al., 2020)** | DL & transfer learning (using pre-trained weights from ImageNet) & specific loss functions and optimization schemes. | CNNs based on ResNet-152 | significance of the fovea and retinal vessels in sex determination | AUC=0.881 (95% CI: 0.877 - 0.885) for images without the fovea<br><br>AUC=0.682 (95% CI: 0.676 - 0.688) for images without the retinal vessels,<br><br>Both compared to an AUC of 0.97 (95% CI: 0.96 - 0.98) for the original images. |
| **Hypertension** | | | | |
| **Poplin et al. (2018) (Poplin et al., 2018)** | Deep neural network | Inception-v3 | Systolic blood pressure prediction | Accuracies=72% (±15 mmHg error margin) |
| | | | Diastolic blood pressure prediction | Accuracies=79% (±10 mmHg error margin) |
| **Dai et al. (2020) (Dai et al., 2020)** | CNN | U-Net | Hypertension prediction | accuracy of 60.94% and AUC of 0.65 |
| **Zhang et al. (2020) (Zhang et al., 2020)** | Transfer Learning and DL | Inception-v3 | Hypertension prediction | 68.8% accuracy and AUC of 0.766 |
| **Atherosclerosis** | | | | |
| **Gao et al. (2021) (Gao et al., 2021)** | CNN-based DL | - | arteriosclerotic retinopathy grading | Accuracy of 65.354% |
| **Chang et al. (2020) (Chang et al., 2020)** | DL | - | Carotid atherosclerosis prediction | Accuracy=0.583 Sensitivity= 0.891 Specificity=0.404 |
| **Rim et al. (2021) (Rim et al., 2021)** | DL | - | RetiCAC prediction (coronary artery calcium scores using retinal photographs) | AUC of 0.742 (95% CI: 0.732–0.753) |
| **Son et al. (2020) (Son et al., 2020)** | DL | Inception-v3 | RetiCAC prediction | AUC=82.3% for detecting absent coronary calcium<br>AUC=83.2% for coronary artery calcium scores greater than 100 |
| **Barriada et al. (2022) (Barriada et al., 2022)** | CNN-based DL | VGG 16 | RetiCAC prediction | Accuracy=0.68 (95% CI: 0.64, 0.72) |
| | | VGG 19 | | Accuracy=0.67(95% CI: 0.59, 0.75) |
| | | ResNet | | Accuracy= 0.64(95% CI: 0.53, 0.75) |
| **Cheung et al. (2021) (Cheung et al., 2021)** | CNN-based DL | - | Associations between measurements of retinal-vessel caliber and cardiovascular disease. | Correlation coefficients of between 0.82 and 0.95 |
| **Left ventricle changes** | | | | |
| **Cheung et al. (2007) (Cheung et al., 2007)** | - | - | Association between concentric LV remodeling and retinal arteriole narrowing | OR: 2.06; 95% CI: 1.57 - 2.70 |
| **Hondur et al. (2020) (Hondur et al., 2020)** | - | - | choroid thickness in patients with CHF | In CHF=231 mm versus in control group=254 mm, p value=0.29) |
| | | | choroidal vascularity index (CVI: choroidal vessels/total choroidal area in patients with CHF | In CHF=0.53 vs in control group=0.60, P value=0.004 |
| | | | the ratio of Haller's layer to total choroidal thickness (Haller/choroid ratio) in patients with CHF | In CHF=0.67 vs in control group=0.64, p value=0.01 |
| **Chandra et al. (2019) (Chandra et al., 2019)** | - | - | Association between central retinal venular equivalent (CRVE), central retinal arteriolar equivalent (CRAE), and incidence of HF | P value < 0.001 |

The choice of the best-performing model or architecture in the context of CVD and its risk factor prediction is contingent upon the specific risk factor and the most pertinent performance metric under consideration. In the domain of CVD risk factor prediction, the MobileNet-V2 model emerges as the top performer for predicting smoking status, while comprehensive summaries of the best-performing architectures for other factors like age and sex are detailed in Table 2. Furthermore, the CNN models presented in Vaghefi et al.'s study perform better for predicting smoking status in the fundus area. Lastly, regarding the prediction of HTN and RetiCAC in atherosclerosis assessment, the Inception-V3 architecture stands out as the leading performer.

**4.6 Databases used for AI model training**

For training AI, a massive amount of training data is required to ensure model accuracy and reliability. High information load forms the foundation on which AI systems improve their learning, enabling them to identify details and patterns in a variety of contexts and promoting generalizability to real-world scenarios. The UK Biobank, which encompasses 22 centers throughout England, Wales, and Scotland, is a pioneering biomedical research initiative launched in the early 2000s to enhance understanding of health conditions and diseases. It has gathered a large clinical, imaging, and genetic dataset from approximately half a million people in the UK. The collected data, including medical records such as retinal images, genetic analyses, lifestyle surveys, and biospecimens, offer an unprecedented opportunity for large-scale, data-centric investigations. This information is accessible to authorized researchers, fostering collaborative efforts that have the potential to drive groundbreaking discoveries in healthcare innovation, thus solidifying the UK Biobank's role in advancing global medical science and improving healthcare outcomes (The UK

Biobank, 2023). Similarly, the Qatar Biobank was intended to be a vital source of information regarding the health-related samples and lifestyle of Qatari nationals. This initiative targets the exploration of risk factors associated with prevalent health issues in the country, i.e., obesity, heart disease, diabetes, and cancers. Of note, The Qatar Biobank is noteworthy for its active involvement in the gathering and use of retinal images for research, which has been crucial in the development of DL for medical applications, such as the detection of CVD and diabetic retinopathy (Al-Absi et al., 2022; The Qatar Biobank, 2023). The EyePACS is an open-access diabetic retinopathy image dataset sourced from a telemedicine service provider that links primary care physicians with ocular specialists in various geographic locations. Streamlined procedures for capturing, transmitting, and evaluating medical images are instrumental in facilitating prompt identification and management of eye disease, and data mining for AI research (Cuadros & Sim, 2004; EyePACS, 2023). The Singapore Epidemiology of Eye Diseases Study is a multi-ethnic longitudinal population-based research investigating age-related eye diseases among Singaporean adults of Malay, Indian, and Chinese descent. Notably, it involves retinal images, making it a valuable resource for researchers in ophthalmology and epidemiology. By providing insights into the incidence, prevalence, risk factors, and novel biomarkers of eye diseases, the study allows researchers to dive deeply into the complexities of eye health within diverse populations (Majithia et al., 2021). The Cardiovascular Risk Prediction with Retinal Data amalgamates clinical data and retinal photographs from diverse data sources, which has enabled productive interdisciplinary collaborations, including cardiovascular risk prediction using fundus images. (Rim et al., 2021). The Age-Related Eye Disease Study 1 database focuses on eye diseases, particularly age-related macular degeneration (ARMD) and cataracts. However, there are potential secondary uses of this database in the field of CVD (Age-Related Eye Disease Studies (AREDS/AREDS2), 2023). The Fundus Image Vessel Segmentation is a specialized dataset tailored for AI development in ophthalmology. Its focus on vessel segmentation of fundus images empowers researchers to explore AI applications for retinal image analysis. It has catalyzed advancements in computer vision and medical imaging, bridging the gap between technology and healthcare (Jin et al., 2022). Initiated in 1975 and funded by the U.S. National Institutes of Health, Structured Analysis of the Retina is a pioneering cornerstone for retinal image analysis research (STructured Analysis of the Retina, 2023). Kaggle often hosts retinal image datasets as part of its data science competitions. Researchers and data scientists can search for relevant competitions and datasets related to retinal image analysis on Kaggle (Kaggle, 2023).

Table 3 provides an overview of the databases used to train AI models in the research.

**Table 3** Summarizes the databases utilized for training AI models in the study.

| Database Name | Description |
|---|---|
| **UK Biobank (The UK Biobank, 2023)** | A large-scale biomedical research initiative in the UK, providing extensive datasets of clinical, imaging, and genetic information from 500,000 UK residents, including retinal images, for large-scale investigations and collaborative research efforts |
| **Qatar Biobank (The Qatar Biobank, 2023)** | A platform focused on gathering health-related samples and lifestyle data from the Qatari population, particularly targeting prevalent health issues like obesity, heart disease, diabetes, and cancers. Includes retinal images for DL research |
| **EyePACS (EyePACS, 2023)** | An open-access diabetic retinopathy image dataset from a telemedicine service provider, facilitates prompt identification and management of eye disease and enables data mining for AI research. |
| **Singapore Epidemiology of Eye Diseases Study (Majithia et al., 2021)** | Longitudinal population-based research investigating age-related eye diseases among Singaporean adults of different ethnicities, providing retinal images and valuable insights into eye health within diverse populations. |
| **Cardiovascular Risk Prediction with Retinal Data (Rim et al., 2021)** | An amalgamation of clinical data and retinal photographs from diverse sources, facilitating interdisciplinary collaborations and cardiovascular risk prediction using fundus images |
| **Age-Related Eye Disease Study 1 (Age-Related Eye Disease Studies (AREDS/AREDS2), 2023)** | Focuses on eye diseases like age-related macular degeneration and cataracts, with potential secondary uses in cardiovascular disease research |
| **Fundus Image Vessel Segmentation (Jin et al., 2022)** | Specialized dataset tailored for AI development in ophthalmology, focusing on vessel segmentation of fundus images to advance retinal image analysis and medical imaging technology |
| **Structured Analysis of the Retina (STructured Analysis of the Retina, 2023)** | A cornerstone database for retinal image analysis research, initiated in 1975 and funded by the U.S. National Institutes of Health, provides structured data for comprehensive analysis |
| **Kaggle (Kaggle, 2023)** | An online platform hosting retinal image datasets as part of data science competitions offers researchers and data scientists access to relevant datasets and opportunities for analysis and innovation. |

### 4.7 AI models utilized in retinal images evaluation

AI methods, particularly those based on deep learning, have shown great potential in retinal image evaluation. They have the ability to identify previously unrecognized imaging features, predict clinical outcomes, and segment anatomic features from ophthalmic imaging. Research shows that a range of models and architectures have been used to study, analyze, and evaluate retinal images and their associated vasculature. Every model has special qualities and uses that contribute to the progress of this field. Figure 3 shows the various architectures for evaluating specific CVD and its risk factors. However, each model has limitations restricting its operations.

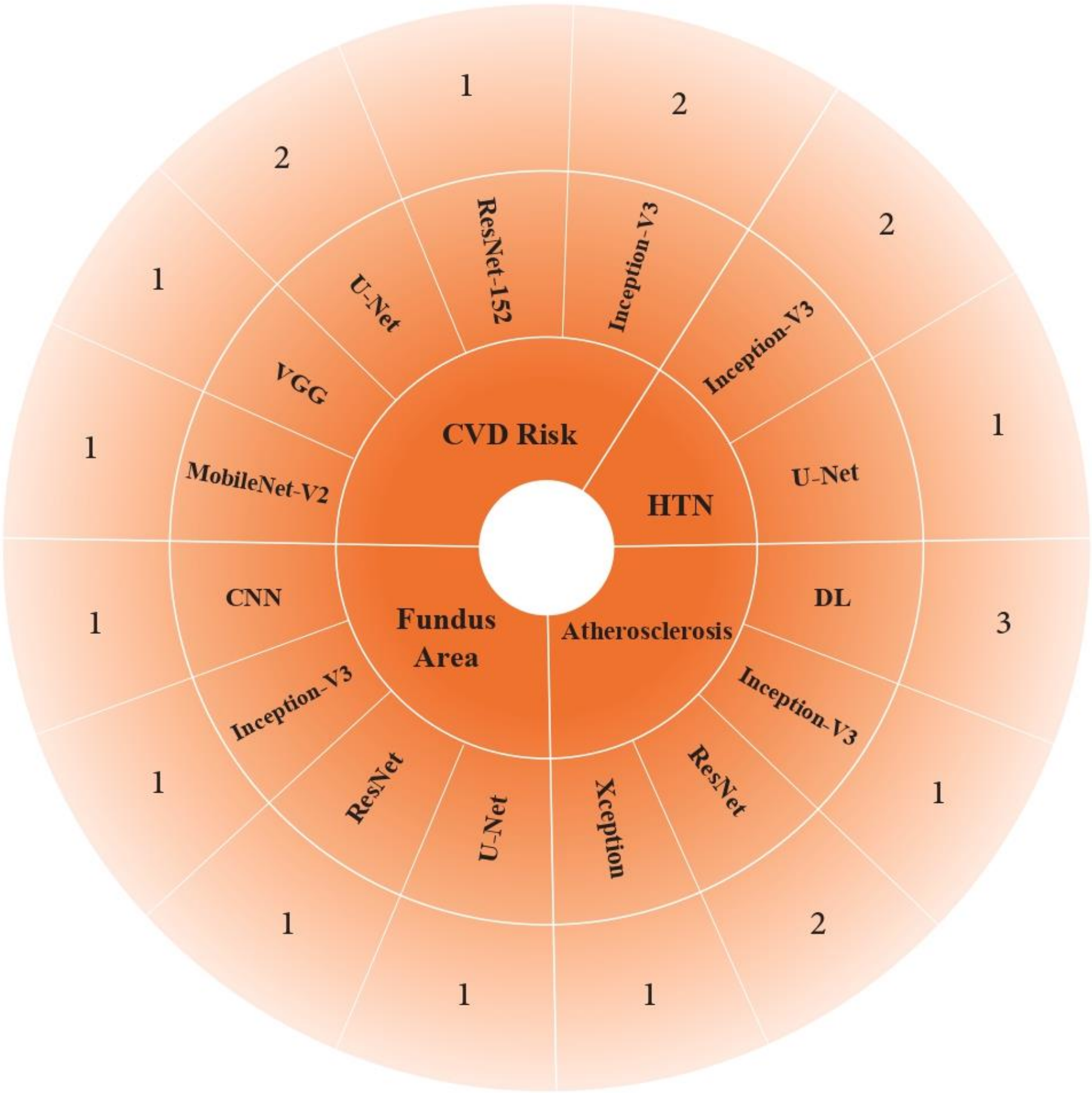

Figure 3 - Radial tree map illustrating the diversity of deep learning models employed across four predominant AI applications in our study. Each sector represents a distinct field of AI application, subdivided to portray the implemented models. The number in each segment reflects the frequency of model usage across different domains, giving information about the distribution and prevalence of deep learning techniques in different applications.

ResNet and Inception-v3 are the two most commonly used advanced CNN architectures, prominently used for image analysis, including retinal images. ResNet is a revolutionary kind of CNN that introduces a novel strategy called "residual learning" to address the "vanishing gradient problem," a common deep learning challenge. Stated differently, ResNet makes use of "residual blocks," which are convolutional layers and skip connections, among other building blocks (Kurama, 2020). These skip connections allow the network to pass gradients without changes during training, which helps prevent the vanishing gradient issue. There are several advantages to this architecture. It enables the creation of more deeper, more flexible, and scalable neural networks. ResNet has, therefore, demonstrated superior performance in various computer vision tasks like image classification, object detection, and segmentation (Reguant et al., 2021). On the other hand, Inception-v3, from the GoogleNet family, is explicitly designed for image classification, focusing on optimizing the network's performance by having complex and multi-level feature extraction mechanisms to handle large-scale image recognition tasks. Inception-v3 is characterized by its 42-layer depth and complexity (Patel, 2020). Its factorization technique breaks down larger convolutions into smaller ones, reducing parameters and improving computational efficiency. It maintains a low top-5 error rate and proves robust for image classification tasks (Yang et al., 2021). For retinal image analysis specifically, both architectures can be applied effectively, with ResNet potentially offering deeper insight due to its ability to train deeper models and Inception-v3 providing an efficient alternative with comparable proficiency (Kurama, 2020).

Some other neural network architectures, such as MobileNet, VGG, and U-Net, are used for retinal image analysis but fundamentally differ in their structures and applications, especially in retinal image analysis (Figure 4). MobileNet is known for its computational efficiency and compactness, making it suitable for devices with limited resources. It is designed for mobile and embedded vision applications, and it uses depth-wise separable convolutions (Sarkar, 2021). VGG-16, Known for its simple and deep architecture, is highly regarded for its image classification performance and is an excellent choice for extracting complex features from images. In contrast, U-Net is mainly used in biomedical image segmentation by using its distinct U-shaped architecture, which allows it to achieve high segmentation precision even when working with a small amount of training data (Hasal et al., 2023; Kugelman et al., 2022). While MobileNet is optimized for quick classifications in constrained environments, VGG-16 is best suited for tasks requiring the extraction of detailed features due to its more profound architecture. U-Net is ideal for segmentation tasks, particularly in biomedical imaging. Selecting one of these models should be based on the project's particular needs regarding application, task orientation, and computational efficiency.

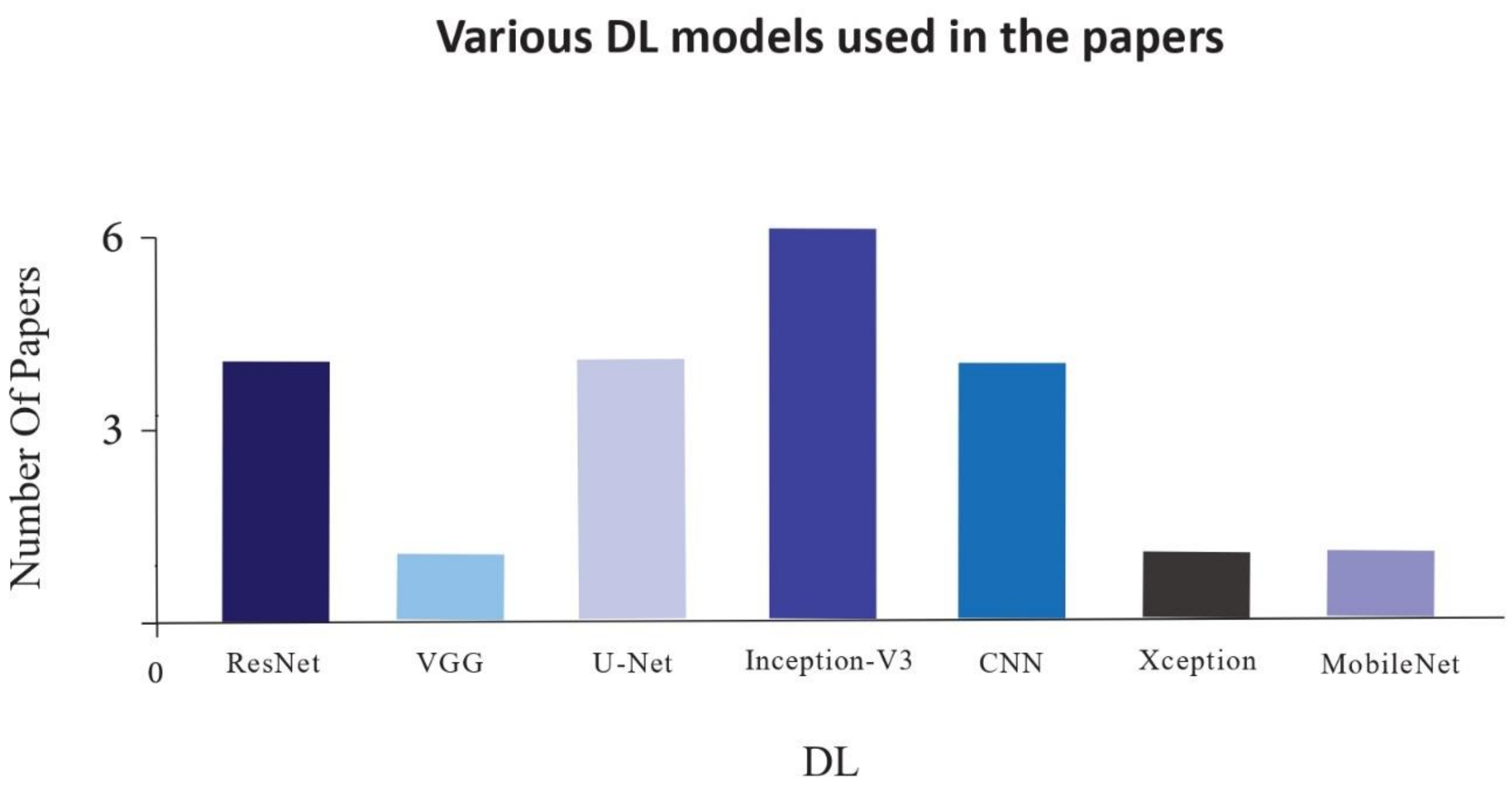

Figure 4 - The stacked bar chart depicts the diverse DL models employed to assess CVD risk factors based on fundus images.

Furthermore, using Bayesian models in retinal image analysis, pre-trained models such as VGG16 and DenseNet201 can be used to classify various retinal diseases. By employing Bayesian statistics to interpret predictive uncertainty, these models enhance robustness and reliability (Subramanian et al., 2022). CNN-based models, on the other hand, are a type of deep learning that uses convolutional layers to effectively process and analyze image data. They work especially well for tasks like image recognition, object detection, segmentation, and various other computer vision tasks (Alyoubi et al., 2020). While CNN models have more simple structures and are naturally optimized for image data, Bayesian models provide valuable insights into prediction reliability, which is important when making medical diagnoses where uncertainty and interpretability are crucial.

**5. Current challenges and prospects**

Overall, while there is a positive perspective regarding the use of AI-based prediction for CVDs, practical implementation still faces several challenges, including:

## 5.1 Uncertainty in AI

In AI, uncertainty refers to the state in which we lack information or find it challenging to accurately predict outcomes due to missing data, noisy inputs, or ambiguous circumstances. When AI systems make decisions, they often encounter uncertainty, which necessitates mechanisms to handle and manage this ambiguity (Nordström, 2022; Seoni et al., 2023). Uncertainties, such as epistemic uncertainty, occur when we lack knowledge about the data or model parameters. Another type of uncertainty stems from data variability or randomness, known as aleatoric uncertainty. Improving decision-making procedures, strengthening AI models against difficulties, and guaranteeing performance in real-world situations all benefit from managing uncertainty. To handle uncertainty in AI systems, methods like quantifying uncertainty and modeling Bayesian inference are essential. These techniques empower them to make reliable decisions in intricate and unpredictable circumstances (Gillies, 2004; Jahmunah et al., 2023).

## 5.2. Concurrent disease

In retinal imaging, the coexistence of multiple retinal diseases may influence the efficacy of AI-driven CVD prediction through varied correlation effects. In clinical environments, it is common for patients to manifest with numerous disorders concurrently. AI systems' inability to make decisions independent of different retinal illnesses may severely limit their application in the general population. Multitask DL algorithms may be more valuable since they can yield many outcomes from a single retinal image, such as CVDs and retinal disorders (Dong et al., 2022).

## 5.3. Data collection and ethical issues

AI methods for diagnosing eye disorders need more high-quality, diverse images for training and performance improvement (Dong et al., 2022). Sharing data from several centers appears to be the best way to increase the amount of input data for network training, but this may need significant financial support. However, increasing the number does not always improve network performance (Ferro Desideri et al., 2022). To effectively predict and classify ocular and systemic diseases using retinal pictures, it is imperative to establish a precise guideline regarding the optimal number of instances required for training. When data is shared across multiple centers, international, national, and state rules, such as patients' privacy, must be considered (Tan & Sun, 2023). Collaboration between regulators, insurance companies, hospital administrators, IT departments, physicians, and patients is essential. AI must be simple and free of significant administrative obstacles to be acceptable, so the rapid distribution of results is critical (Miller, 2022). Additionally, obtaining informed consent is necessary for the practice of medicine, but incorporating AI into diagnostics poses challenges. Healthcare professionals need to inform patients about the use of AI in their treatment and ensure they are aware of the implications of AI deployment (Astromskė et al., 2021).

AI-driven illness prognostication poses challenges, especially ethical dilemmas. Patients may be reluctant to rely entirely on the opinions of computers. Physicians might be wary of AI decisions because there is not a clear chain of accountability. Before AI systems are used exclusively for treatment decision-making, clear guidelines must be established (Aung et al., 2021; Basu et al., 2020).

Disparities in healthcare outcomes could arise from AI algorithms' unintentional propagation of bias. it is crucial to understand that AI algorithms are only as unbiased as the data they are trained on. biases present in the training data may be learned and maintained by the AI algorithm, leading to unfair or biased outcomes (Nazer et al., 2023). To address this, several strategies can be employed, including collecting diverse and representative datasets, implementing bias-aware algorithm design techniques, continuous monitoring and evaluation of AI systems, and promoting inclusivity and teamwork in development teams. These strategies aim to mitigate biases, ensure fairness and equity, improve transparency, and promote trust among stakeholders (Mittermaier et al., 2023).

In order to improve accuracy and AUC values in this field, future research can investigate architectures such as ResNet, DenseNet, or custom-designed models and incorporate larger datasets from a variety of fundus images from different populations, races, or societies, employing advanced training techniques and conducting thorough external validations to improve accuracy and AUC values in this field.

### 5.4. Data quality and interoperability

Implementating AI in healthcare faces several technical challenges, including data quality, interoperability, and security. Since healthcare data is frequently erroneous, inconsistent, and incomplete and can, therefore, introduce biases and errors, data quality is a crucial consideration in AI applications in medicine. Interoperability challenges are primarily centered around the integration of diverse data sources (Torab-Miandoab et al., 2023). It is challenging to extract and integrate data from multiple sources in many healthcare applications because they lack the necessary data export functions or application programming interfaces (APIs). To overcome these obstacles, interoperability standards and frameworks must be established (Arora et al., 2023).

### 5.5. Data Privacy

Strong solutions and increased attention are required to address the ethical and legal complexities surrounding data privacy in the context of AI-driven healthcare. The relationship between patients and doctors is significantly impacted by these implications (Jafarizadeh et al., 2024). Concerns about privacy and data ownership are frequent when it comes to digital systems, making patient-informed consent necessary. Sensitive information is protected by strong measures like data encryption and established consent protocols (Abdullah et al., 2021).

### 5.6. High dimensionality of data

The high dimensionality of data in medical datasets poses computational challenges due to the many variables involved. This complexity can hinder data processing and analysis. To address this issue, techniques such as relevant variable selection and data preprocessing play a crucial role in reducing dimensionality, improving computational efficiency, and enhancing the accuracy of analytical models (Kalian et al., 2023).

### 5.7. Interpretability of AI models

Interpretability of AI models is crucial in gaining trust and acceptance, especially in medical applications where understanding the decision-making process is essential. Visualization techniques play a key role in enhancing the interpretability of AI models, allowing stakeholders to grasp the inner workings and reasoning behind the model's predictions. By visualizing the model's outputs using methods like heat maps or saliency maps, stakeholders can gain insights into the regions or features in the input data that influence the model's decision-making process, thus improving transparency and trust (Ennab & McHeick, 2022).

### 5.8. Financial support

The progress of AI research, development, and application is significantly influenced by financial support. Funding and grants are crucial for the advancement of AI technologies and the application of research findings. Working with government agencies, cost-sharing programs, and industry-supporting organizations is necessary to promote sustainable growth in the AI landscape. By working together with these groups and putting cost-sharing plans in place, researchers and innovators can secure the funding needed to advance AI projects and ensure their sustainability across a range of fields

### 5.9. Multidisciplinary collaboration

The advancement of effective and morally sound AI solutions depends heavily on multidisciplinary collaboration. By bringing together professionals from diverse fields, including data science, artificial intelligence, and medicine, organizations can leverage various perspectives and expertise to drive innovation and address complex challenges. Centralized platforms and multidisciplinary teams are crucial tools for facilitating efficient communication, exchanging knowledge, and encouraging collaboration among specialists from various fields. This cooperative approach improves the development of AI solutions and guarantees that these technologies are developed and implemented responsibly, taking ethical considerations and societal impact into account (Dwivedi et al., 2021).

### 5.10. Adherence to rules and regulations

Ensuring adherence to rules and regulations is essential when integrating AI into healthcare systems to maintain patient safety and data privacy. Compliance with laws like the US Health Insurance Portability and Accountability Act (HIPAA) is critical for upholding ethical standards. Legal experts provide valuable guidance in navigating the complex regulatory landscape surrounding AI in healthcare. Harmonized frameworks tailored to medical AI applications can further ensure responsible implementation. By following established regulations, healthcare providers and technology developers uphold ethical standards and promote the secure and ethical use of AI in healthcare (Pesapane et al., 2021).

### 5.11. Generalizability

Enhancing the analysis of AI models in CVD prediction requires consideration of various factors affecting their generalizability. Challenges arise due to patient diversity, encompassing variations in demographics, genetics, and lifestyle factors. Data bias, stemming from underrepresentation of certain demographic groups, can hinder model generalizability and lead to predictive inaccuracies across populations. Moreover, the diverse clinical presentations of CVDs necessitate robust validation across demographic and clinical subgroups. Ethnic and socioeconomic disparities in healthcare access and quality further impact data availability and model generalizability. Cultural differences in healthcare practices underscore the importance of culturally sensitive approaches for effective CVD prediction. To address these challenges and pave the way for more inclusive and impactful AI models in CVD prediction, tailored approaches and strategies are needed. This may involve adopting localized AI systems designed for the unique contexts of LMICs, as well as collaborative initiatives sensitive to local contexts, benefiting diverse patient populations (Gulati et al., 2022; Singh et al., 2022; Wang et al., 2022). The challenges of AI in the context of CVD risk assessment are succinctly summarized in Figure 5**.**

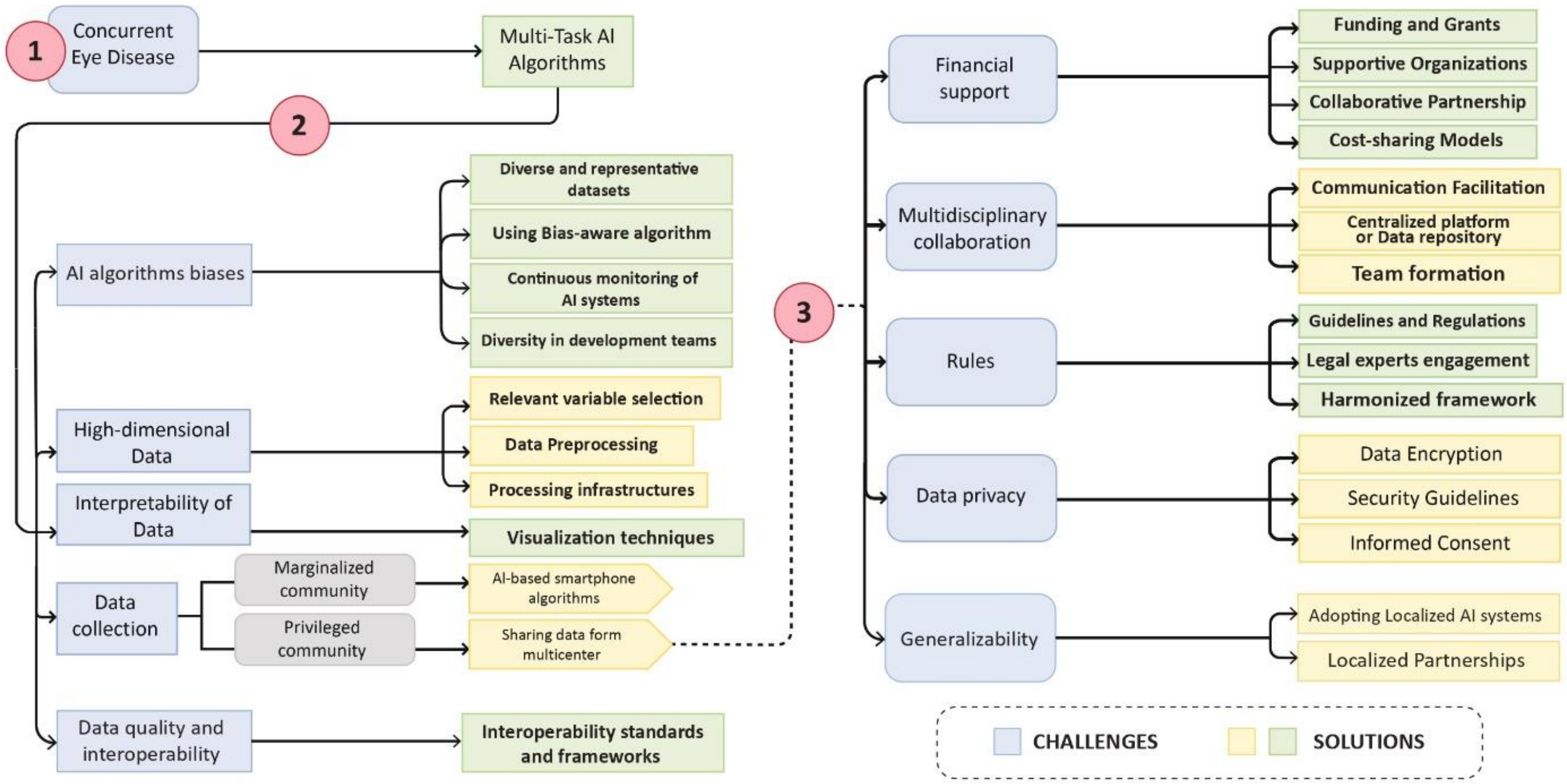

Figure 5 - An overview of the challenges in applying AI to CVD risk assessment and corresponding solutions to overcome these challenges. On the left side, the challenges posed by the utilization of multi-task AI algorithms and their solutions are represented. On the right side, there are additional challenges related to data sharing among multiple centers and their corresponding solutions.

## 6. New horizon

AI algorithms can analyze large datasets containing individual patient's genetic information, lifestyle habits, environmental factors, and medical history. By recognizing patterns and correlations in this data, AI can predict an individual's risk of developing CVD. However, it is important to remember that the successful application of AI in personalized medicine requires overcoming several challenges, such as ensuring data privacy, managing the high dimensionality of data, and improving the interpretability of AI models. Furthermore, the potential DL should not be conflated with the proficiency of a specialist. While these DL networks can excel when tasked with specific objectives, they cannot replace the expert judgment of trained retina specialists. That said, the advent of AI-based smartphone algorithms enables even those without expert training to capture and analyze images that could be significant in predicting CVD risks. With the integration of AI-based analytical tools on smartphones and through teleophthalmology, the prospect of screening for CVD risks among the general population and marginalized communities appears increasingly feasible. It would be beneficial to survey patients and healthcare providers to confirm the technology's practical efficacy in primary healthcare settings. There is great potential for improving our knowledge of the risk factors for cardiovascular disease through the combination of fundus eye images and human genetic profile analysis. Zekavat et al. (Zekavat et al., 2022) performed a Genome-Wide Association Study (GWAS) on retinal vascular density and fractal dimension. The study used a sample of 38,932 unrelated. The researchers analyzed 15,580,782 variants with a minor allele frequency greater than 0.001. Through their analysis, they identified 13 loci significantly associated with retinal vascular density and seven loci significantly associated with retinal fractal dimension at a genome-wide level. Encouraging further research endeavors in this domain will not only facilitate the creation of a specialized gene mapping framework but also enable a more comprehensive assessment of cardiovascular disease susceptibility through precise genetic matching. Therefore, researchers must embark on further studies in this field, as they have the potential to uncover invaluable insights that may revolutionize the way we approach cardiovascular disease prevention and management.

Explainable AI (XAI) allows humans to understand and trust AI algorithm output. It tries to make AI models more visible and interpretable, revealing how they make decisions. Explainable AI tackles the "black-box" character of many machine learning models, which are difficult to comprehend (Hulsen, 2023). It clarifies AI model predictions, helping people trust the algorithm's output. XAI can help identify and mitigate errors in healthcare processes, explain recommendations, and ensure safer patient care. XAI models can analyze medical data, such as retinal images, electronic health records, and genetic information, to identify early signs of diseases. XAI can improve images by assisting radiologists in interpreting medical images. It can highlight regions of interest and explain findings, reducing diagnostic errors (Loh et al., 2022).

## 7. Conclusion

Researchers are looking for a new and precise series of patterns to address how to detect the prognosis and severity of diseases as the global population ages and life expectancy rises. Almost always, CVDs are one of the top three reasons for mortality worldwide. As a result, early diagnosis and risk stratification significantly reduce healthcare costs and raise society's standard of health. The changes in the retina's structure and blood vessels have reflected changes in certain parts of the cardiovascular system, which has medical significance. Recently, how we live and practice medicine has altered due to the increased usage of AI systems. The sector will continue to change over the next few years, but more needs to be done to accelerate the adoption of AI in healthcare systems. The idea of a machine that can scan various CVDs only from a single retinal image could become a reality as scientists develop novel methods with greater learning capacities.

**Funding information**

None.

**Authors' contributions**

AJ and RA conceptualized the manuscript. MA, AJ, NS, SP, HA, and AGA conducted a literature review and drafted the manuscript. AJ, MA, KP, RA, RST, and URA provided comments and revised the manuscript. NS and AJ created and designed all the figures. All authors read and approved the final version of the manuscript. It should be noted that AJ and RA are co-corresponding authors of this manuscript.

**Acknowledgments**

We are grateful to Grammarly and ref-n-write for their services for editing and paraphrasing.